\documentclass[twocolumn,pre,showpacs,showkeys]{revtex4-1}
\usepackage{graphicx}
\usepackage{soul}
\usepackage{color}
\usepackage{amsmath}
\usepackage{pgf}
\usepackage[greek,english]{babel}
\usepackage{subcaption}
\usepackage{caption}
\usepackage{amssymb}

\begin{document}

\title{Dynamical properties of neuromorphic Josephson junctions} 

\author{D. Chalkiadakis}
\affiliation{Department of Physics, University of Crete, 71003 Herakleio, Greece
}
\author{J. Hizanidis} \email[Correspondence email address: ]{hizanidis@physics.uoc.gr}

\affiliation{Department of Physics, University of Crete, 71003 Herakleio, Greece
}
\affiliation{Institute of Applied and Computational Mathematics,
Foundation for Research and Technology-Hellas,
70013 Herakleio, Greece}
\date{\today}

\begin{abstract} 
Neuromorphic computing exploits the dynamical analogy   between many physical systems and neuron biophysics. 
Superconductor systems, in particular, are excellent candidates for neuromorphic devices due to their capacity to operate in great speeds and with low energy dissipation compared to their silicon counterparts.
In this study we revisit a prior work on Josephson Junction-based ``neurons" in order to identify the exact dynamical mechanisms underlying the system's neuron-like properties and reveal new complex behaviors which are relevant for neurocomputation and the design of
superconducting neuromorphic devices. Our work lies at the intersection of superconducting physics and theoretical neuroscience, both viewed under a common framework, that of nonlinear dynamics theory.  

\end{abstract}



\maketitle
\section{Introduction}

Neuromorphic computing is a rapidly advancing field that uses neuroscience-inspired concepts in order to implement circuits of physical neurons. The ultimate goal of neuromorphic computing is the development of powerful algorithms and high-speed, energy-efficient hardware for information processing and the potential acquirement of insight into cognition (for a recent review see~\cite{MAR20} and references within). The motivation behind the attempt to mimic the brain is its 
extremely impressive capabilities and advantages as a computing device, in terms of storage, processing speed, memory and energy consumption.

The reason for its outstanding performance lies in the brain's complexity, specifically the fact that it is dynamic and reconfigurable (due to plasticity), it provides large interconnectivity, it is stochastic, and exhibits interesting nonlinear phenomena like synchronization and chaos,  
to mention only a few of the brain's characteristics~\cite{NIC91,VRE98}. The latter, in particular, have inspired nonlinear dynamics based computing, which utilizes the many different intrinsic behaviors of a nonlinear dynamical system for performing different types of computation~\cite{HOP99,KIA15}.

Neuromorphic computing exploits the dynamical and especially the nonlinear-dynamical 
analogy between many physical systems and neuron biophysics. Various implementations of neuromorphic systems have been proposed, namely CMOS (complementary metal oxide semiconductor) and memristor devices~\cite{MIL20,HOE20}, photonic networks~\cite{SHA21}, spintronic nanondevices~\cite{GRO20} and superconductor systems. In light of the recent advances in new materials and hardware, the development of
increasingly efficient neuromorphic devices 
is challenging yet promising (for a detailed comparison between the aforementioned different approaches see~\cite{MAR20}).

Superconductor-based neuromorphic systems are particularly advantageous since they are very fast, with operation speeds close to THz, and most importantly, 
present very low or no power dissipation, even when cryogenic cooling is taken into account. Over the last years, there has been a significant increase in the number of implementations of neuromorphic devices using superconducting elements such as superconducting
quantum interference devices (SQUIDS)~\cite{MIZ94}, quantum-phase slip junctions~\cite{CHE18}, superconducting nanowires~\cite{TOO19,LOM21}, and Josephson Junctions (JJs)~\cite{Segall10,Segall17,SEG14,SCH20,BRA16}. 
The latter produce the so-called single flux quantum (SFQ) pulse~\cite{LIK1986} which is qualitatively very similar to the action potential that is fired by real neurons when the membrane potential exceeds its threshold.

Most works on JJ neuromorphic devices involve circuit simulations and theoretical modelling (for a recent review see ~\cite{SCH22}).
However, several experimental implementations demonstrate that such devices can be indeed fabricated and easily engineered for neuromorphic applications. More specifically, in 
~\cite{Segall17} a circuit of two mutually coupled excitatory neurons was studied both numerically and experimentally. Each neuron was realized using Josephson junctions, a Josephson transmission line acted as the axon, and the synapse was modelled by a SQUID
similarly to prior works~\cite{MIZ94}. It was found that the neurons are either desynchronized or synchronized in an in-phase or antiphase state, and that
the tuning of the delay and strength of the SQUID
synapses can switch the system back and forth in a phase-flip bifurcation~\cite{SEG14}.

The building block of Josephson Junction neuromorphic circuits is the single JJ neuron model, which was developed over a decade ago in ~\cite{Segall10}. There it was shown that  the JJ neuron is capable of reproducing many characteristic behaviors of biological neurons such as as action potentials, refractory periods, and firing thresholds. In the present work, we revisit Ref.~\cite{Segall10} and perform an extended study on the complex behavior of single Josephson Junction neurons in order to shed light on new dynamics which can further inform the design of devices, and discuss the associated neurocomputational properties this system is capable of presenting. Our work lies at the intersection of superconducting physics and theoretical neuroscience, viewed under the framework of nonlinear dynamics theory. 

The paper is organized as follows: in Sec.~\ref{sec:JJ Neuron Model} we derive the Josephson junction neuron model and describe the mechanism for the production of the ``action potential". In Sec.~\ref{sec:Dynamics} the system's complexity is explored through bifurcation analysis and focus is given on its excitable behavior (Sec.~\ref{sec:Excitability}) and the chaotic and multistable dynamics it presents (Sec.~\ref{sec:Chaos}). Finally, in Sec.~\ref{sec:Neurocomputational properties}, we identify the neuronal properties emulated by the model and stress their significance in terms of neural computation. We summarize our results in Sec.~\ref{sec:Conclusions}.

\section{Josephson Junction Neuron Model}
\label{sec:JJ Neuron Model}
\begin{figure}[h!]
    \centering
    \begin{subfigure}{0.35\linewidth}
        \includegraphics[width=\linewidth]{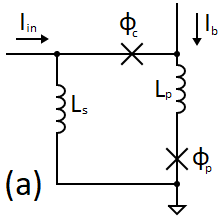}
        \phantomsubcaption{}
        \label{fig:JJneuron}    
    \end{subfigure}
    \begin{subfigure}{0.60\linewidth}
        \includegraphics[width=\linewidth]{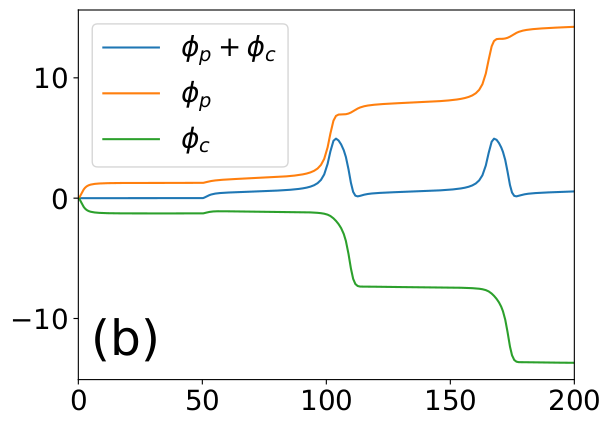}
        \phantomsubcaption{}
        \label{fig:spike_formation}    
    \end{subfigure}
    
    \caption{(a) Circuit diagram for the Josephson Junction neuron. (b) The voltage of the membrane is emulated by the quantity $\phi_p+\phi_c$. Increasing the current from zero to $i_{in}=0.22$ at $t=50$ forces the JJ Neuron to spike. System parameters: $\Lambda_p=0.5$, $\Lambda_s=0.5$, $\lambda=0.1$, $\Gamma=1.5$ and $i_b = 1.909$. Initial conditions: $(0,0,0,0)$.}
    \label{fig:Intro}
\end{figure}

As implied by its name, the Josephson Junction neuron (JJ neuron) involves two Josephson junctions, in a loop, as shown in the  circuit depicted in Fig.~\ref{fig:JJneuron} (where JJs are marked with an $\mathrm{X}$). A JJ is a nonlinear superconducting element made by two superconductors connected through a ``weak link" such as an insulator. The fundamental properties of JJs have been
established long ago~\cite{JOS62} and have been exploited in
numerous applications in superconducting electronics, sensors, and high frequency devices ever since. Each superconductor of the JJ can be described by a single macroscopic wavefunction with a corresponding phase, and the difference between these two phases is the so-called Josephson phase, denoted by $\phi$. 

In an ideal JJ, the (super)current through the JJ and the voltage across the JJ are related through the celebrated Josephson relations: $I=I_\text{cr} \sin(\phi)$ and $V = (\hbar/2e)d\phi/d\tau$
where $I_\text{cr}$ is a critical current above which the voltage develops, $\tau$ denotes the time, $e$ is the electron charge and $\hbar$ is the Planck's constant.
Within the framework of the  Resistively and Capacitively Shunted Junction (RCSJ) model \cite{LIK1986}, the current flowing through the junction is given by Kirchhoff's law and contains contributions from a displacement current and an ordinary current, represented by a capacitor $C$ and a resistor $R$, respectively:
\begin{equation}
    \frac{\hbar C}{2e}\frac{d^2\phi}{d\tau^2} + \frac{\hbar}{2eR}\frac{d\phi}{d\tau}+I_\text{cr}\sin{\phi}=I.
    \label{JJ_equation}
\end{equation}
The mechanical analog of the JJ is the damped pendulum driven by a constant torque. 
Depending on the initial conditions, the strength of the drive and the damping, the solution of such a system may involve static tilting, whirling modes, or a combination of the two~\cite{STR94}.
In the Josephson junction, the ``whirling" of the phase, when the applied current exceeds a critical value, creates a magnetic flux pulse~\cite{LIK1986}. This single flux quantum forms the basis for the pulse produced by the JJ neuron model, which is qualitatively very similar to the action potential that occurs in real neurons when the membrane potential exceeds its threshold. 

A schematic plot of the JJ Neuron is shown in the circuit of Fig.~\ref{fig:JJneuron}. The two (identical) Josephson junctions connected in a superconducting loop are called ``pulse" and ``control" junctions, and are denoted by the subscripts ``p'' and ``c'', respectively~\cite{Segall10}. By simplifying Eq.~\ref{JJ_equation} using the following normalizations: $t^2 =\tau^2 (2eI_\text{cr}/\hbar C) $, $\Gamma^2 = \hbar/(2eI_\text{cr}R^2C)$, $i = I/I_\text{cr}$, and by direct application of Kirchhoff’s laws, we obtain the dimensionless equations for the phases of the JJ neuron circuit:

\begin{align}
    \ddot{\phi_p}+\Gamma \dot{\phi_p}+\sin{\phi_p}&=- \lambda(\phi_p+\phi_c)+\Lambda_s i_\text{in}+ (1-\Lambda_p)i_b\label{ddot_phip},\\
    \ddot{\phi_c}+\Gamma \dot{\phi_c}+\sin{\phi_c}&= - \lambda(\phi_p+\phi_c)+\Lambda_s i_\text{in} -\Lambda_p i_b\label{ddot_phic},
\end{align}
where the dot notation refers to differentiation with respect to $t$, $\Lambda_s$ and $\Lambda_p$ are the inductances $L_p$ and $L_s$, respectively, scaled by their sum $L_\text{tot}$, the currents $i_b$ and $i_\text{in}$ are scaled by the critical current $I_\text{cr}$, and finally $\lambda=\hbar/2eL_\text{tot}I_\text{cr}$ is the coupling parameter. The bias current $i_b$ provides necessary amounts of energy to both junctions, while the current $i_\text{in}$ emulates the incoming postsynaptic current received by the neuron.

For appropriate parameter values the magnetic flux in the JJ neuron $\lambda (\phi_p+\phi_c)$ emulates the voltage difference across the neuronal membrane. In Fig.~\ref{fig:spike_formation} we visualise $\phi_p+\phi_c$, omitting $\lambda$ because it is just a scaling factor, in order to demonstrate the generation of the action potential in the JJ neuron. The stimulus $i_\text{in}$ is sufficiently strong after $t>50$, so that it forces the phase $\phi_p$ to increase abruptly (blue curve in Fig.~\ref{fig:spike_formation}). The coupling between $\phi_p$ and $\phi_c$, regulated by $\lambda$, causes the opposite reaction for the phase $\phi_c$ (red curve in Fig.~\ref{fig:spike_formation}). The combined effect of the two phases results in the creation of a pulse (green-coloured in Fig.~\ref{fig:spike_formation}) which is qualitatively very similar to the action potential of a real neuron.

The analogy between the JJ neuron and the biological one also extends to the voltage across the pulse and control junctions: $u_p$ and $u_c$ correspond to the ionic currents flowing in real neurons, $Na^+$ and $K^+$, respectively, which underlie the generation of the action potential~\cite{DAY01}. After the JJ neuron fires, the phase $\phi_p$ slowly starts to build-up again, $\phi_c$ reacts accordingly (as described previously), and this results in a refractory period-like behavior, before the next spike occurs (Fig.~\ref{fig:spike_formation}). For further details on the creation of the JJ Neuron action potential one may refer to \cite{Segall10}.

In this study the parameters $\lambda=0.1,\Lambda_p = \Lambda_s =0.5$ are fixed as in previous works~\cite{Segall10,SEG14,Segall17}. Similarly, the bias current is kept constant at a typically used value $i_\text{b}=1.909$.
In the Appendix, we investigate the role of $i_\text{b}$ and explain why values close to but lower than $2$ should be used. In the following sections we explore the role of $\Gamma$, which remains unaltered after the fabrication of the Josephson junction, and that of $i_\text{in}$, which in principle is tunable.

\section{Dynamics of the JJ neuron}
\label{sec:Dynamics}
The JJ neuron model described in the previous section
reproduces many characteristic properties of biological neurons such as action potentials and firing thresholds~\cite{Segall10}. 
In this work we aim to study these properties in a more systematic way, in terms of bifurcation analysis, and explore further the complexity of the system's dynamics and the corresponding neuronal behaviors they relate to.

\subsection{Excitability and Bistability}
\label{sec:Excitability}

One of the basic dynamical properties of a neuron which is related to the transition between firing and resting states is excitability, i.~e. the ability of a neuron to realize a large amplitude change in its membrane voltage, in response to an external stimulus which is above a certain threshold. Excitability is fundamental beyond neurons, in many physical systems, such as semiconductor structures~\cite{HIZ06,ORT21} and lasers~\cite{DUB99}.
There are typically two types of excitability depending on the relationship between the firing frequency and the applied stimulus intensity~\cite{HOD48}. The generated action potential in Type I neurons increases with increasing the applied stimulus, whereas type II neurons exhibit a finite nonzero frequency as periodic firing begins. 

The JJ neuron is capable of demonstrating both types of excitability depending on the system parameter values as reported in \cite{Segall10}. However, a full bifurcation analysis of the system's excitability is missing. In the present study, we  perform a bifurcation analysis and continuation in the relevant parameter space in order to identify, in detail, the regions of spiking/resting and the transitions between them. 
\begin{figure*}[!htp]
    \centering
    \includegraphics[width=0.9\linewidth]{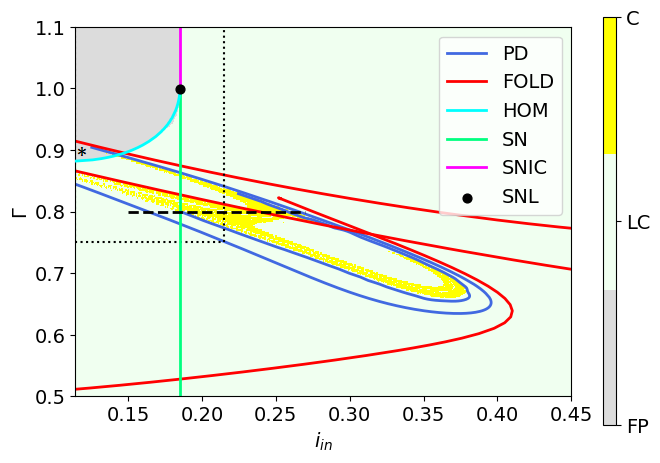}
    \caption{Chaotic (C), periodic (LC), and resting (FP) dynamics over the parameter plane $(i_\text{in},\Gamma)$ according to the Lyapunov spectrum. The following bifurcation lines are superimposed: period doubling (PD), fold of cycles (FOLD), homoclinic (HOM), saddle node of fixed points (SN), saddle node on invariant circle (SNIC), and saddle-node loop (SNL). The square defined by the black dotted lines marks the parameter subspace of Fig.~\ref{fig:RouteToChaos} and the  black dashed horizontal line marks the parameter value range of Fig \ref{fig:Cod2_Bif}. The starred area contains fixed points as well as very small periodic and chaotic windows which are not easily visible. Other parameter values: $\lambda=0.1,\Lambda_p = \Lambda_s =0.5, i_\text{b}=1.909$. Different initial conditions were considered, while, in case of coexistence we choose to visualise the dynamics as such: C over LC, and LC over FP.}
    \label{fig:Attractors_bifurcations}
\end{figure*}

The transition from resting to spiking occurs through the collision of a stable with an unstable fixed point. Equations \ref{phi_p_star_equation}-\ref{phi_c_star_equation} of the Appendix provide the equilibria and can be used to detect at which $i_\text{in}$ values they annihilate. The corresponding bifurcation lines that separate the regions of spiking and resting are depicted in Fig.~\ref{fig:Attractors_bifurcations} and are analysed in the following.

\begin{figure*}[!htp]
    \centering
    \begin{subfigure}{0.46\linewidth}
        \includegraphics[width=\linewidth]{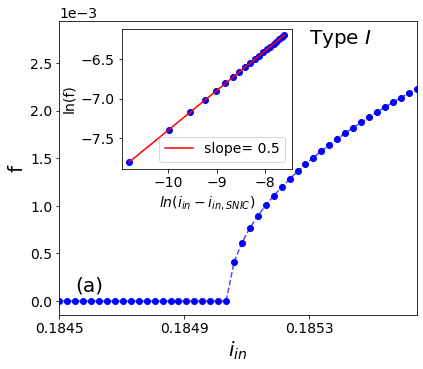}\hfil
        \phantomsubcaption{}
        \label{fig:SNIC}    
    \end{subfigure}
    \begin{subfigure}{0.475\linewidth}
        \includegraphics[width=\linewidth]{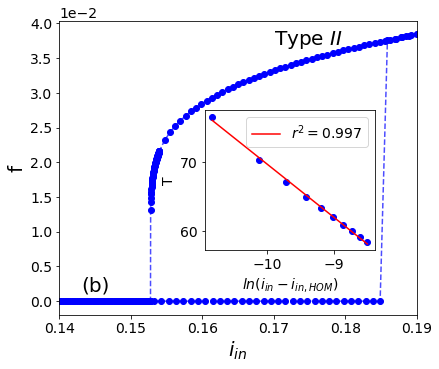}
        \phantomsubcaption{}
        \label{fig:homοclinic}
    \end{subfigure}
    
    \caption{Frequency of the limit cycle over the stimulus $i_\text{in}$: (a) For $\Gamma = 1.5$ spiking occurs through a SNIC bifurcation. (b) For $\Gamma = 0.9$ the curve forms a hysteresis loop between $i_\text{in,HOM}$ and $i_\text{in,SN}$ where the neuron either spikes or rests depending on the initial conditions. The inset graphs depict that sample points, near the HOM and SNIC bifurcations obey the corresponding scaling laws $T_\text{HOM} \sim O(\ln{[i_\text{in}-i_\text{in,HOM}]})$ and $f_\text{SNIC} \sim O(\sqrt{i_\text{in}-i_\text{in,SNIC}})$ respectively. Other parameters: $\Lambda_p=0.5$, $\Lambda_s=0.5$, $\lambda=0.1$ and $i_b = 1.909$.}
    \label{F-I_curves}
\end{figure*}

For $\Gamma>1$ the transition occurs through a saddle node on an invariant circle (SNIC) bifurcation, marked by the purple line in Fig.~\ref{fig:Attractors_bifurcations}. At the bifurcation point $i_\text{in,SNIC}= 0.185$, a stable limit cycle is born whose frequency follows the scaling law: $f \sim O(\sqrt{i_\text{in}-i_\text{in,SNIC}})$. The square root law is verified by the 0.5 slope in the semi-logarithmic plot in the inset of Fig.~\ref{fig:SNIC}, where the frequency of the limit cycle is plotted as a function of the stimulus $i_\text{in}$. Exactly at the bifurcation point, the period of the limit cycle is infinite, therefore this bifurcation is also know as saddle-node infinite period bifurcation (SNIPER) and it characterizes neurons of excitability type I~\cite{RIN98}.

On the other hand, for $\Gamma<1$, the resting state disappears at a stimulus value $i_\text{in,SN}=i_\text{in,SNIC}$ through a saddle-node bifurcation (SN), in this case \emph{off limit cycle}, marked with a light green line in Fig.~\ref{fig:Attractors_bifurcations}, forcing the trajectories to follow an already existing limit cycle of nonzero frequency. The aforementioned limit cycle is born through a homoclinic (HOM) bifurcation (cyan line in Fig.~\ref{fig:Attractors_bifurcations}), at a stimulus $i_\text{in,HOM}<i_\text{in,SN}$. The HOM bifurcation was detected through its characteristic scaling law of the period of the LC near the bifurcation point, which should follow: $T \sim O(\ln{[i_\text{in}-i_\text{in,HOM}]})$. Indeed, the inset of Fig.~\ref{fig:homοclinic}, where the limit cycle frequency is plotted over $i_\text{in}$, verifies the above relation with an R-squared value of $r^2 = 0.997$. For a fixed $\Gamma$ value, the JJ neuron is bistable for $i_\text{in} \in (i_\text{in,HOM},i_\text{in,SN})$, since a limit cycle coexists with a stable equilibrium. This accounts for class II excitability.

As already mentioned, these bifurcations were detected and identified through the F-I curves which are visualised in Fig.~\ref{F-I_curves}. More specifically, we first fixed $\Gamma$ and then used the following protocol: For each value of the current, the frequency was calculated and the last variable values of the trajectory were used as the initial conditions for the next simulation. In this way the trajectories start at the vicinity of the attractor which was detected in the previous run. In Fig. \ref{fig:homοclinic} we first increased the current from 0.14 to 0.19 and then moved backwards, investigating the current values near the homoclinic bifurcation.

It can be easily shown that the system is unable to undergo Hopf bifurcations, which are also known to be related with class II neural excitability. During a Hopf bifurcation, the conjugate pair of two eigenvalues of a fixed point must cross the imaginary axis~\cite{KUZ98}, which is impossible in this system. This stems from Eqs.~\ref{eig_1}--\ref{eig_4} of the Appendix which reveal that when the system has complex eigenvalues, their real part is $\operatorname{Re}(e_i)=-\Gamma/2$, which is always a negative quantity and never passes through zero.

At $\Gamma=1.0$ the two bifurcation points $i_\text{in,HOM}$ and $i_\text{in,SN}$ coincide.
The process where a saddle-node and a homoclinic bifurcation coalesce forming a SNIC bifurcation is called a saddle node separatrix loop (SNL) or saddle-node homoclinic orbit bifurcation, marked by a full dot in Fig.~\ref{fig:Attractors_bifurcations}. The SNL bifurcation is common to all class I excitable neurons~\cite{Sch21} and has also been found in the single Josephson junction model (see \cite{SCH87} and references within). To summarize, around the SNL bifurcation, the line $i_\text{in}=0.185$ separates spiking from resting behavior. In addition, next to the resting regime, there 
exists a bistable portion of the parameter space which is bound from above by the homoclinic bifurcation line.
As we move toward lower $\Gamma$ values in the parameter space, we encounter additional bifurcations, shown in Fig.~\ref{fig:Attractors_bifurcations} which lead to more complex dynamics including multistability and chaotic spiking, as we will see in the next section.

\subsection{Chaotic dynamics and multistability}
\label{sec:Chaos}

From a mathematical point of view, the JJ neuron is a four-dimensional nonlinear dynamical system and, therefore, capable of presenting a plethora of complex phenomena. In this section, we will focus on the chaotic and multistable dynamics exhibited by the system. The chaotic regimes are detected according to the Lyapunov spectrum, which was extracted using the ``Dynamical Systems'' Julia package \cite{DynamicalSystems}. The Lyapunov spectrum consists of four Lyapunov exponents $L_i$, sorted in descending order, with their sum following $\sum_i^{4} L_i = \det J = -2\Gamma<0$, where $J$ is the Jacobian provided in the Appendix (Eq. \ref{Jacobian}).

Since $L_4$ is always negative, the three largest Lyapunov exponents are sufficient for characterizing the dynamics of the JJ neuron.
Figure~\ref{fig:Attractors_bifurcations} demonstrates the different dynamical regimes according to the Lyapunov spectrum.
More specifically, for (I) $L_\text{1,2,3}<0$ the system's solution is a fixed point (FP, light gray), for (II) $L_\text{1}=0$, $L_\text{2,3}<0$ the system's solution is a limit cycle (LC, pale green), while for (III) $L_\text{1}>0$, $L_\text{2}=0$, $L_\text{3}<0$ the system exhibits chaotic behavior (C, yellow). For the generation of Fig.~\ref{fig:Attractors_bifurcations} we considered different initial conditions whereas, in case of coexistence of two or more attractors, we visualise the attractor according to the following order: chaos over limit cycle and limit cycle over equilibrium.
In addition, two different types of bifurcation lines are superimposed, namely period doubling (PD) and fold of cycles (FOLD), marked with blue and red color, respectively. The bifurcation lines have been
obtained using a very powerful software tool that executes a root-finding algorithm for continuation of periodic solutions~\cite{ENG02}. 


The bifurcation structure of the system is very intricate: The fold and PD bifurcation lines intersect the homoclinic and saddle-node line discussed in Sec.~\ref{sec:Excitability}, creating thus two smaller areas, one ``triangular``-shaped corresponding to bistability and the starred area in Fig.~\ref{fig:Attractors_bifurcations} which mostly contains a single fixed point and some very small windows where the FP coexists with a limit cycle or chaos. 
Moving toward smaller values of $\Gamma$, the system's dynamics becomes much more complex and involves multiple periodic solutions which are created and destroyed through fold bifurcations of cycles, as well as oscillatory and chaotic states coexisting with the stable equilibria. From Fig.~\ref{fig:Attractors_bifurcations} it is evident that the transition from periodic to chaotic motion takes place through a period-doubling route to chaos~\cite{STR94}. 
This can be illustrated more clearly if we focus on the cross section of the parameter space for $\Gamma=0.8$, $i_\text{in} \in [0.15,0.27]$, marked by the dotted black line in Fig.~\ref{fig:Attractors_bifurcations}.

\begin{figure*}[!htp]
    \centering
    \includegraphics[width=\linewidth]{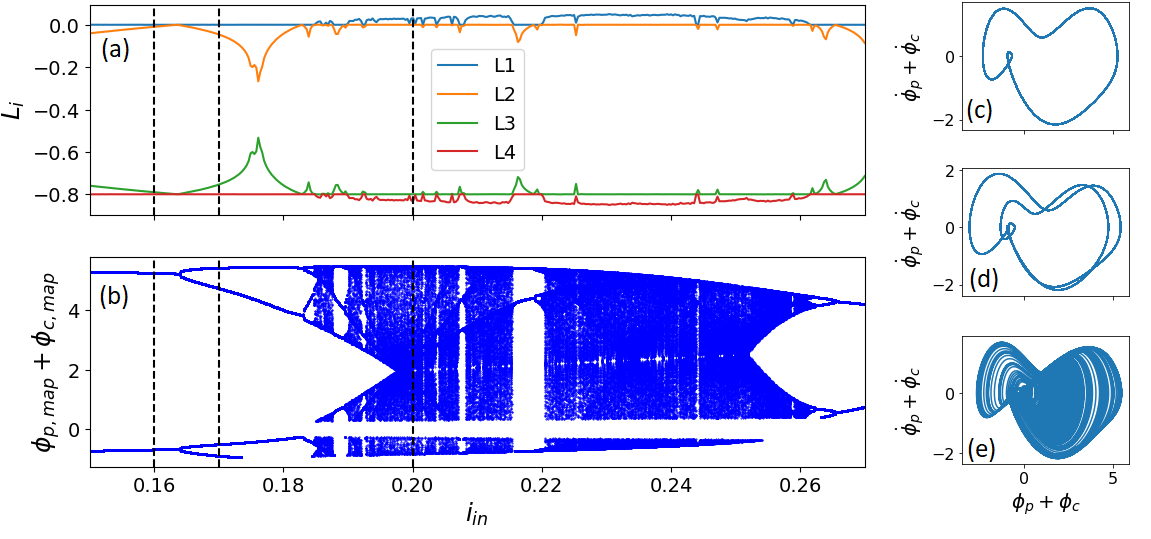}
    \caption{Route to chaos for $\Gamma=0.8$: (a) Lyapunov spectrum, (b) Orbit diagram of the Poincar\'e map whose surface of section is the plane $\dot{\phi_p}+\dot{\phi_c}=0$. Dashed lines depict the values of $i_\text{in}$ which were chosen for the visualisation of the phase portraits. More specifically: (c) for $i_\text{in}=0.16$, (d) for $i_\text{in}=0.17$ and (e) for $i_\text{in}=0.20$. Other parameter values: $\lambda=0.1$, $\Lambda_p = \Lambda_s =0.5$, $i_b=1.909$. Each time the system is initialised close to the attractor detected in the previous run, as in Fig. \ref{F-I_curves}.}
    \label{fig:RouteToChaos}
\end{figure*}

The blow-up of this region is shown in Fig.~\ref{fig:RouteToChaos}, where the Lyapunov spectrum as a function of $i_\text{in}$ is plotted. For $0.15<i_\text{in}<0.1632$ it holds that $L_\text{1}=0$, $L_\text{2,3}<0$ and the dynamics is, therefore, periodic. At $i_\text{in}=0.1632$ the two largest Lyapunov exponents become zero $L_1=L_2=0$ and the first period doubling bifurcation occurs. This is followed by a cascade of period doubling bifurcations which lead to chaos, where $L_\text{1}>0$, $L_\text{2}=0$, $L_\text{3}<0$. Note that, for simplicity, in Fig.~\ref{fig:Attractors_bifurcations}, we have only plotted the outer PD line that includes the first period-doubling bifurcation. 

The route to chaos is also reflected in the corresponding Poincar\'e map, shown in Fig.~\ref{fig:RouteToChaos}, where we store the value $\phi_\text{p,map}+\phi_\text{c,map}$ each time the trajectory crosses the plane $\dot{\phi_p}+\dot{\phi_c}=0$ and $\ddot{\phi_p}+\ddot{\phi_c}<0$. This particular selection of the variables and plane of intersection is not arbitrary as the stored quantities are the local maxima of the JJ neuron response. The first simulation of Fig. \ref{fig:RouteToChaos}, that is for $i_{in}=0.15$,  was initialised at $(\phi_\text{p,0},\omega_\text{p,0},\phi_\text{c,0},\omega_\text{c,0})=(0,20.0,0,0)$, same as in Fig. \ref{fig:Attractors_bifurcations}. The following runs on the other hand, were initialised with the last variable values of the previous simulation. Thanks to this protocol, we are able to follow the evolution of the attractor which becomes chaotic without falling into the resting state, or other LCs. Comparing the Lyapunov spectrum of Fig.~\ref{fig:RouteToChaos}a with the Poincar\'e map of Fig.~\ref{fig:RouteToChaos}b it is clear that when $L_1=L_2=0$ the branches of the map split in two, which is the signature of the period doubling bifurcation. 
This transition to chaos is also visualized in Figs.~\ref{fig:RouteToChaos}(c-d) where the phase portraits in the $(\phi_p+\phi_c,\dot{\phi_p}+\dot{\phi_c})$ plane are shown, for values of the control parameter $i_\text{in}$ marked by the vertical dashed lines in Fig.~\ref{fig:RouteToChaos}(a-b). At $i_\text{in}=0.16$ the system has a period-1 solution (Fig.~\ref{fig:RouteToChaos}a), which doubles its period after the first PD bifurcation (Fig.~\ref{fig:RouteToChaos}b), and consequently undergoes a cascade of period-doublings  before entering chaos (Fig.~\ref{fig:RouteToChaos}c).

\begin{figure}[htp]
    \centering
    \includegraphics[width=\linewidth]{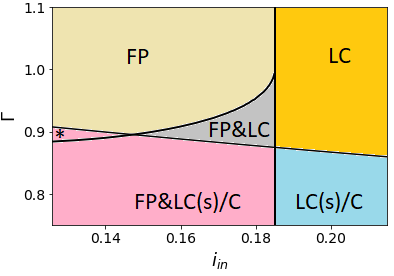}
    \caption{ Different dynamical regimes in the parameter space $(i_{in},\Gamma)$. The detected dynamics are fixed points (FP), limit cycles (LC), chaos (C), and coexistence thereof, in various combinations. The starred area corresponds to the same area in \ref{fig:Attractors_bifurcations} and it mostly contains exclusively a FP, and some very small windows of coexistence of a FP and a LC or C. System parameters: $\Lambda_p=0.5$, $\Lambda_s=0.5$, $\lambda=0.1$ and $i_b = 1.909$.}
    \label{fig:Cod2_Bif}
\end{figure}

In order to have an overview of the menagerie of behaviors exhibited by the JJ neuron, we have created a mapping of all the different dynamical regimes analyzed above, shown in Fig.~\ref{fig:Cod2_Bif}. Previous works on JJ neurons focused on regimes where the sole attractor is a periodic orbit~\cite{Segall10}.
However, the system is capable of presenting a plethora of additional
dynamics and the knowledge of its full behavior is useful for the design
of experiments based on JJ neurons and particularly their exploitation with relevance to neurocomputation.

\section{Neurocomputational properties of the JJ Neuron}
\label{sec:Neurocomputational properties}

The dynamical behavior described in the previous section determines the neurocomputational properties of a JJ neuron. The JJ neuron is known to be capable of reproducing many characteristic behaviors of biological neurons~\cite{Segall10}. In this section we extend these findings by identifying additional neuronal properties emulated by the JJ model and stressing their significance in terms of  neural computation.

In Sec.~\ref{sec:Excitability} we confirmed via bifurcation analysis that the JJ neuron is capable of mimicking neurons of both classes I and II of excitability. Both classes of excitability have been observed in biological experiments, for instance in pyramid neurons in the hippocampus and interneurons in the neocortical region, respectively~\cite{PRE08,TIK15}, among others. Differences in excitability result in differences in spike initiation, which in turn has implications for essential biological functions of the brain such as information encoding and processing~\cite{IZH07,RIN98}. Moreover, different classes of neuronal excitability can affect the collective behavior of the nervous system, particularly the phenomenon of synchronization is shown to be achieved more easily in a neuronal network with class II neurons rather than that with neurons of class I~\cite{HAN95}.

Regarding the JJ neuron, the key element in the dynamics relating to both classes of excitability is the SNL co-dimension 2 bifurcation depicted in Fig.~\ref{fig:Attractors_bifurcations}. This bifurcation is found in other famous neuronal models such as the Morris-Lecar and Wilson-Cowan models~\cite{IZH00}. Moreover, it is linked to other neurocomputational properties, namely the existence of a well-defined threshold, all-or-none behavior, and spike latency \cite{IZH00}. The latter property is related to the ``bottleneck" created at the SNIC and SN bifurcations and refers to the existence of significant delays, which can reach up to a second in real neurons, in the production of the first spike when the stimulus is barely greater than the threshold~\cite{IZH04}.

Another interesting feature related to the SNL bifurcation is that one can potentially switch between the two classes of excitability, simply by tuning $\Gamma$ and keeping all other system parameters fixed. The transition between classes of neuronal excitability has been observed in biological experiments~\cite{PRE08} and recently it was reported that such transitions may be induced by autapses~\cite{ZHA17}, i.~e. synapses from a neuron onto itself via closed loops.


The neurons we have encountered can be in a quiescent state or they can fire, either regularly or chaotically. When a neuron alternates between these two states periodically it is said to be bursting. 
In autonomous bursting, that is for constant stimulus, there should be generally an additional variable with a slower timescale than those participating in the spiking, which is responsible for turning off and on the generation of the action potentials \cite{IZH07}.
For this reason, even though 4 dimensional systems such as the JJ neuron are in principle capable of displaying bursting, we have not detected this kind of behavior in our model. The existing model is capable of emulating another type of bursting which is induced by noise rather than some intrinsic mechanism \cite{Sch21}. We should mention at this point that the latter has been achieved in networks of globally coupled mixed populations of oscillatory and excitable Josephson Junctions ~\cite{HENS15,MIS21}.

\begin{figure}[!htp]
    \centering
    \begin{subfigure}{0.9\linewidth}
        \includegraphics[width=\linewidth]{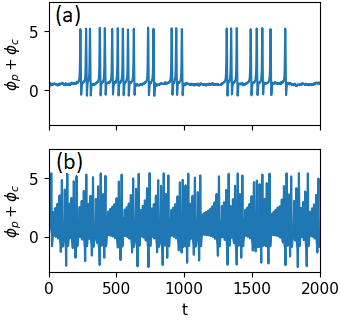}
        \phantomsubcaption{}
        \label{fig:NoiseBursting}
    \end{subfigure}
    
    \begin{subfigure}{0\linewidth}
        \phantomsubcaption{}
        \label{fig:Chaotic}
    \end{subfigure}
    \caption{Different neurocomputational properties of a JJ neuron: (a) Noise induced bursting for $\Gamma = 0.95 $ and stimulus $i_\text{in}=0.182+0.02 \xi(t)$, where $\xi(t)$ is Gaussian white noise,  (b) Chaotic spiking for $\Gamma = 0.8 $ and $i_{in}=0.2$. Other parameter values: $\lambda=0.1,\Lambda_p = \Lambda_s =0.5$, $i_b=1.909$.}
    \label{fig:chaos_bistability}
\end{figure}
Let us now assume a JJ neuron which is in the bistable regime where resting and spiking states coexist. In order to model the variation of the stimulus due to fluctuations we incorporate an additive Gaussian white noise term $\xi(t)$ with amplitude of $0.02$.  The stochastic differential equations were integrated with Milstein's method \cite{TOR14}. The addition of noise helps the system alternate between spiking and quiescence, resulting in a bursting-like behavior as shown in the time-series depicted in Fig.~\ref{fig:NoiseBursting}. 
Both bistability and bursting behaviors have been found in recordings of biological neurons~\cite{IZH07}, while the latter is also considered to be linked to a distinct mode of neuronal signalling~\cite{KRA04}.

In the same figure, we have also plotted the case of chaotic dynamics as we analyzed in Sec.~\ref{sec:Chaos}. Figure \ref{fig:Chaotic} displays a typical example of chaotic firing of the JJ neuron for $\Gamma=0.8$ and $i_\text{in}=0.2$. 
Chaotic behavior in neurons has been extensively studied both in real recordings of neuronal activity \cite{HIR12} and in mathematical models \cite{INN07}, and has been found to be very crucial in terms of cognitive functions. In particular,  due to their information carrying capacity, chaotic  attractors may serve as
as information processors and cognitive devices~\cite{NIC91,NIC15}. 
Moreover, chaos and bifurcations can be exploited for nonlinear dynamics based computing~\cite{HOP99,KIA15}.

Very recently, in \cite{Hochstetter2021}, authors discovered that an artificial network of metallic nanowires with synapse-like memristive junctions can be tuned to respond in a brain-like way when electrically stimulated. More specifically, they found that by keeping this network of nanowires in a brain-like state at the edge of chaos, it performed tasks at an optimal level. These results suggest that neuromorphic devices can be tuned into regimes with different, brain-like collective dynamics, which may be exploited to optimise information processing.

\begin{figure}[!htp]
    \centering
    \begin{subfigure}{\linewidth}
        \includegraphics[width=\linewidth]{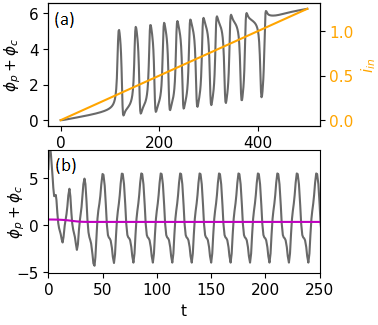}
        \phantomsubcaption{}
        \label{fig:PeriodicFp}
    \end{subfigure}
    \begin{subfigure}{0\linewidth}
        \phantomsubcaption{}
        \label{fig:NonSpike}
    \end{subfigure}
    \caption{Behaviors beyond biological relevance: (a) Periodic behavior of equilibria with increasing $i_\text{in}$ for $\Gamma=1.5$, (b) Non spike-like limit cycle depicted in gray for $\Gamma = 0.7$ and $i_\text{in} = 0.15$ and a coexisting stable fixed point depicted in purple. Other parameter values: $\lambda=0.1,\Lambda_p = \Lambda_s =0.5$, $i_b=1.909$.}
    \label{fig:PeriodicFP_NonSpike}
\end{figure}

Finally, we would like to address some behaviors beyond biological relevance found in the JJ neuron which are particularly interesting. First of all, the equilibria appear and disappear periodically with respect to the stimulus $i_\text{in}$ and independently of $\Gamma$, as shown in the Fig.~\ref{fig:fp_N_over_i} of the Appendix. In this work, we have investigated a certain regime of the parameters, that is for $i_b = 1.909$ and $i_\text{in} \in [0.0,0.2]$, where increasing the stimulus results in the disappearance of the stable equilibria. Larger input values however, such as the ones depicted in Fig.~\ref{fig:PeriodicFp}, may force a spiking neuron to rest, which is not biologically plausible according to our knowledge. Furthermore, Fig.~\ref{fig:Attractors_bifurcations} reveals that there are one or more periodic and even chaotic attractors which coexist with the resting states, especially for $\Gamma<1$. In some cases, these attractors are created or destroyed by means of fold bifurcation of limit cycles or period doubling bifurcations. For example, Fig.~\ref{fig:NonSpike} shows a periodic solution which does not have the familiar spike-like form. 

\section{Conclusions}
\label{sec:Conclusions}
In summary, Josephson Junction neurons are excellent candidates for playing an important part in neuromorphic computing due to their capacity to operate in great speeds and with low energy dissipation compared to their silicon counterparts. For this reason, their dynamical behavior must be fully understood and compared with that of biological neurons.

In this study we confirmed the existence of a saddle node loop separatrix (SNL) bifurcation which was detected in the relevant parameter plane. The SNL bifurcation has been found in mathematical models of neurons and is linked with many neurocomputational properties such as: excitability of class I or II, existence of a well-defined threshold, all-or-none behavior spike latency, and bistability.
All these properties have been identified in biological experiments
and are linked to essential computational functions of the brain.

Apart from the SNL bifurcation, the model was also found to exhibit chaotic and multistable dynamics. By means of Lyapunov exponent calculations and bifurcation analysis, we have identified that this is achieved through a period doubling route to chaos mechanism. Chaotic behavior in real neurons has been verified in the lab in numerous experiments. This type of behavior is of particular importance, as the brain is thought to operate best at the edge of chaos, i.~e. at a critical transition point between randomness and order. 
The JJ neuron also exhibits noise-induced bursting, while autonomous bursting could possibly be achieved by coupling the bias current $i_\text{b}$ with some other variable of the system, for example the voltage of the ``p'' junction, $\dot{\phi_p}$. 
A complete mapping of all the possible dynamics presented by the JJ neuron has been created and can be used to inform 
the design of relevant experiments.

Finally, we also report on other properties of the JJ neuron which are beyond biological relevance such as non spike-like periodic trajectories and a periodic dependence of the equilibria on the input stimulus. Further investigations of the JJ neuron could involve the implementation of a synapse and the study of the coupled system in light of our new findings, or more interestingly the modelling of excitatory and inhibitory neural autapses.

\subsection*{Acknowledgement}
This work was supported by the General Secretariat for
Research and Technology (GSRT) and the Hellenic Foundation for
Research and Innovation (HFRI) (Code No. 203).
D. C. would like to thank Joniald Shena for valuable discussions.

\appendix*
\section{}
\label{sec:appendix}
In the following we derive expressions for the fixed points 
of the JJ neuron system and perform a linear stability analysis 
in order to determine their stability. Defining $\dot\phi_p = \omega_p $ and $\dot\phi_c = \omega_c$, the original system of Eqs. \ref{ddot_phip}-\ref{ddot_phic} is transformed to:
\begin{align}
        \dot\phi_p &= \omega_p,  \\
        \dot\omega_p &= - \Gamma\omega_p-\sin{\phi_p} -\lambda(\phi_c+\phi_p)+\Lambda_si_\text{in}+(1-\Lambda_p)i_b, \\
        \dot\phi_c &= \omega_c, \\
        \dot\omega_c &=- \Gamma\omega_c-\sin{\phi_c} -\lambda(\phi_c+\phi_p)+\Lambda_si_\text{in}-\Lambda_pi_b.
\end{align}
In this way, the evolution of the system can be visualised as a trajectory in the phase plane $(\phi_p,\omega_p,\phi_c,\omega_c)$. Then, the equilibria of the system are $(\phi_p^{\star},0,\phi_c^{\star},0)$ where $\phi_p^{\star},\phi_c^{\star}$ are provided by solving the following Eqs:

\begin{align}
    &\sin{\phi_p^{\star}}-\sin{(- \frac{\sin{\phi_p^{\star}}}{\lambda}-\phi_p^{\star}+\frac{\Lambda_s i_\text{in}+(1- \Lambda_p)i_b}{\lambda})} = i_b\label{phi_p_star_equation},\\
    &\phi_c^{\star} = -\frac{\sin{\phi_p^{\star}}}{\lambda}-\phi_p^{\star}+\frac{\Lambda_s i_\text{in}+(1- \Lambda_p)i_b}{\lambda}\label{phi_c_star_equation}.
\end{align}

\begin{figure}[htp]
    \centering
    \includegraphics[width=\linewidth]{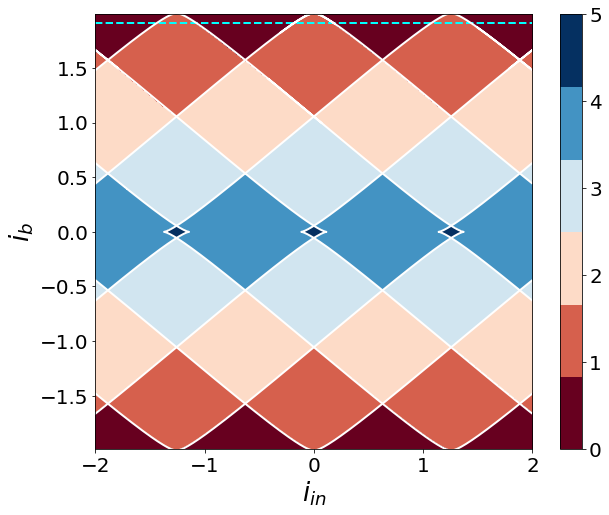}
    \caption{Number of stable fixed points provided by Eq. \ref{phi_p_star_equation} for $i_\text{in} \in [-2,2]$ and $i_b \in (-2,2)$. White lines mark the saddle-node bifurcation. Other parameter values: $\lambda=0.1,\Lambda_p = \Lambda_s =0.5$. Independent of $\Gamma$. The cyan dotted line depicts $i_\text{b}=1.909$.}
    \label{fig:fp_N_over_i}
\end{figure}

The next step is to calculate the stability of the equilibria. Thus, the Jacobian was found as:
\begin{equation}
    J = \begin{bmatrix}0 & 1 & 0 &0\\
    -\cos{\phi_p}-\lambda & -\Gamma &-\lambda &0\\
    0 & 0 & 0 &1 \\
    -\lambda & 0& -\cos{\phi_c} - \lambda & -\Gamma 
    \end{bmatrix}.
    \label{Jacobian}
\end{equation}
The characteristic equation is given by:
\begin{equation}
    \begin{split}
        e^4+2e^3\Gamma +e^2[\cos{\phi_p}+\cos{\phi_c}+2\lambda+\Gamma^2]+e\Gamma[\cos{\phi_p}\\+\cos{\phi_c}+2\lambda]+\lambda(\cos{\phi_p}+\cos{\phi_c})+\cos{\phi_p}\cos{\phi_c},
        \label{characteristic}
    \end{split}
\end{equation}
while the roots of Eq. \ref{characteristic} provide the eigenvalues:

\begin{align}
    e_1 &= \frac{1}{2}(-\sqrt{-A+B}-\Gamma) \label{eig_1},\\
    e_2 &= \frac{1}{2}(\sqrt{-A+B}-\Gamma) \label{eig_2},\\
    e_3 &= \frac{1}{2}(-\sqrt{A+B}-\Gamma) \label{eig_3},\\
    e_4 &= \frac{1}{2}(\sqrt{A+B}-\Gamma) \label{eig_4},    
\end{align}
where

\begin{align}
    A &= 2\sqrt{(\cos{\phi_p}-\cos{\phi_c})^2+4\lambda^2}>0,\\
    B &= -2(\cos{\phi_p}+\cos{\phi_c}+2\lambda)+\Gamma^2.
\end{align}

Notice that $\Gamma$ does not affect the position of the fixed point since it is not contained in the equations \ref{phi_c_star_equation}-\ref{phi_p_star_equation}. Moreover, using Eqs. \ref{eig_1}-\ref{eig_4} one can show that this is also the case for the sign of the real part of the eigenvalues. Thus, the stability of the equilibria is also independent of $\Gamma$. On the other hand, the value of $\Gamma$ affects whether the fixed point is a focus, i.e contains complex eigenvalues, or a node. 

Figure \ref{fig:fp_N_over_i} shows the number of stable fixed points in the $(i_\text{in}, i_\text{b})$ parameter plane, while the white lines mark the saddle-node bifurcation lines through which they lose their stability. We observe that the absolute value of $i_\text{b}$ affects the number of fixed points more decisively than the stimulus $i_\text{in}$. When $|i_\text{b}|$ is small, there are many fixed points while when $|i_\text{b}|>2$ there are no equilibria. A typical neuron is expected to rest until the stimulus exceeds a certain threshold value where firing starts. That is why $i_b=1.909$ was chosen, in order to ensure spiking behavior. Note, finally, that the graph of Fig.~\ref{fig:fp_N_over_i} is periodic over the stimulus $i_\text{in}$, and the borders between different colors mark the annihilation/generation of fixed points. 

Finally, notice that Eqs. \ref{eig_1}-\ref{eig_4} reveal the reason why the system does not exhibit Hopf bifurcations. A fixed point can have a pair of complex conjugate eigenvalues only when the radicand is less than zero. During the Hopf bifurcation, the real part of those eigenvalues must cross the real axis. In this case  $\Gamma$ must flips sign, which is impossible since it corresponds to a positive damping coefficient.

\end{document}